\documentclass[sigconf,screen,authorversion]{acmart}

\usepackage{booktabs} 
\usepackage{tabularx}

\usepackage{multicol}
\usepackage{multirow}
\usepackage{colortbl}
\usepackage{xcolor}

\usepackage{listings}

\definecolor{brightgray}{gray}{0.9}
\definecolor{lightgray}{gray}{0.8}
\definecolor{darkgray}{gray}{0.5}
\definecolor{brightblue}{rgb}{0.8,0.85,1}
\definecolor{lightblue}{rgb}{0.7,0.75,1}
\definecolor{darkblue}{RGB}{20,86,227}
\definecolor{brightgreen}{rgb}{0.8,1,0.8}
\definecolor{cAgile}{RGB}{189,215,238}
\definecolor{cTraditional}{RGB}{255,166,134}
\definecolor{cBoth}{RGB}{164,165,169}

\AtBeginDocument{%
  \providecommand\BibTeX{{%
    \normalfont B\kern-0.5em{\scshape i\kern-0.25em b}\kern-0.8em\TeX}}}

\copyrightyear{2024} 
\acmYear{2024} 
\setcopyright{acmlicensed}\acmConference[ICSSP '24]{International Conference on Software and Systems Processes}{September 4--6, 2024}{M\"unchen, Germany}
\acmBooktitle{International Conference on Software and Systems Processes (ICSSP '24), September 4--6, 2024, M\"unchen, Germany}
\acmDOI{10.1145/3666015.3666021}
\acmISBN{979-8-4007-0991-3/24/09}

\begin{document}

\title{Software Process as a Service: Towards A Software Process Ecosystem}


\author{Oliver Greulich}
\affiliation{%
  \institution{Technische Universität Clausthal}
  \department{Institute for Software and Systems Engineering}
  \city{Clausthal-Zellerfeld}
  \country{Germany}}
  \email{oliver.greulich@tu-clausthal.de}

  \author{Christoph Knieke}
  \orcid{0009-0006-8018-2351}
\affiliation{%
  \institution{Technische Universität Clausthal}
  \department{Institute for Software and Systems Engineering}
  \city{Clausthal-Zellerfeld}
  \country{Germany}}
  \email{christoph.knieke@tu-clausthal.de}

\author{Bassel Rafie}
\affiliation{%
  \institution{Technische Universität Clausthal}
  \department{Institute for Software and Systems Engineering}
  \city{Clausthal-Zellerfeld}
  \country{Germany}}
  \email{bassel.rafie@tu-clausthal.de}

\author{Andreas Rausch}
\orcid{0000-0002-6850-6409}
\affiliation{%
  \institution{Technische Universität Clausthal}
  \department{Institute for Software and Systems Engineering}
  \city{Clausthal-Zellerfeld}
  \country{Germany}}
  \email{andreas.rausch@tu-clausthal.de}
  
\author{Marco Kuhrmann}
\orcid{0000-0001-6101-8931}
\affiliation{%
	\institution{Reutlingen University}
	\department{Faculty of Informatics~|~Herman Hollerith Center}
	\streetaddress{Danziger Str. 6}
	\postcode{71034}
	\city{B\"oblingen}
	\country{Germany}
}
\email{kuhrmann@acm.org}

\renewcommand{\shortauthors}{Greulich and Knieke, et al.}

\begin{abstract} 
In large-scale projects operated in regulated environments, standard development processes are employed to meet strict compliance demands.
Since such processes are usually complex, providing process users with access to their required process, which should be tailored to a project's needs is a challenging task that requires proper tool support.
In this paper, we present a process ecosystem in which software processes are provided as web-based services. We outline the general idea, describe the modeling approach, and we illustrate the concept's realization using a proof-of-concept case based on a large software process line that is mandatory to use for IT projects in the German public sector. The suitability is evaluated with three experts that valued the improved accessibly and usability of the process and the end-user support tool.
\end{abstract}




\maketitle

\section{Introduction}
Traditionally, software processes are devised as a mixture of documentation parts and supporting tools. For instance, the Eclipse Process Framework (EPF; \cite{eclipse-process-framework}) served as editor to write the process documentation and, at the same time, acted as framework to add further tools \cite{ruiz2012uses,kuhrmann2013systematic}. Such tools, however, required complex installation procedures. For instance, for the German V-Modell~XT \cite{DBLP:conf/ispw/KuhrmannNR05,kuhrmann2011survey}, platform-specific installation procedures were devised to install the required tools---just to generate process documentation or project plan templates. Meanwhile, this heavy-weight client-side tool infrastructure has been replaced by web-based solutions. Specifically, agile methods like Scrum\footnote{\url{https://www.scrum.org}} or XP\footnote{\url{http://www.extremeprogramming.org}} quite often are documented using compact websites. Even complex processes follow the web-based approach, e.g., processes documented using process management solutions such as MethodPark's Stages\footnote{\url{https://www.methodpark.com/stages.html}}. That is, even complex processes have entered the web-age, yet, quite often as process documentation tool only with third-party tools being configured from the process model in the background. Those processes, however, remain individual process exemplars, i.e., comprehensive process management, such as process lines, and evolution-related features are scarcely to find.

\paragraph{Problem Statement and Objectives}
In large-scale and regulated environments, standard or \emph{reference processes} are used due to strict compliance requirements. For instance, in the automotive sector, processes have to comply with the ISO/IEC~26262 \cite{iso26262-1}. Likewise, in the European aerospace sector, the compliance with the ECSS-standards\footnote{\url{https://ecss.nl/standards}} is mandatory. However, such reference processes are generic and require context-specific refinements and embodiment, which leads to so-called \emph{Software Process Lines} (SPrL; \cite{10.1007/11608035_9,de2014software}) that underly a strict management including evolution and compliance processes, and rules to derive compliant process variants. This paper addresses the problem of providing users of complex process models with access to their required process variant within a software process line, specifically tailored to project's needs using modern web technologies. Our objective is thus \emph{to design a process ecosystem in which software processes and software process variants are provided as web-based services.}

\paragraph{Contribution}
The paper's contribution is two-fold: (i) we propose a concept and a reference implementation of a \emph{Software Process as a Service} including a description of the transition of a metamodel-based SPrL into a web-environment. (ii) We evaluate the current concept and implementation by analyzing  the infrastructure implementation and the implications the concepts and implementations have for tool vendors and process end users.

\paragraph{Outline}
The rest of the paper is organized as follows: In Section~\ref{sec:BGAndRW}, we provide background information about the used technologies and related work. Section~\ref{sec:SPrLEcosystem} introduces the software process ecosystem including the modeling approach, rules, technical designs and realizations. Section~\ref{sec:CaseStudy} presents a proof-of-concept validation, before we conclude the paper in Section~\ref{sec:Conclusion}.

\section{Background and Related Work}
\label{sec:BGAndRW}
In this section, we provide an overview of the background and related work. In Section~\ref{sec:BGAndRW:BG}, we provide the background in terms of the OpenAPI Specification (OAS) used to design and realize the process ecosystem. Section~\ref{sec:BGAndRW:RW} discusses related work.

\subsection{Technical Background}
\label{sec:BGAndRW:BG}
The OpenAPI Specification (OAS; \cite{OpenAPI}) defines a standard, programming language-agnostic interface description for HTTP-based APIs, which enable both humans and computers to explore and comprehend the functionalities of a service without the need for having access to source code, extra documentation, or scrutiny of network traffic. When accurately specified through OpenAPI, a consumer can grasp and engage with the remote service with minimal implementation logic required. The API can be specified in \emph{Json} or \emph{Yaml} as machine-readable specification. With the help of \emph{SwaggerUI}, a specification can be visualized and used interactively. OpenAPI and the complementing software infrastructure lay the technical foundation for the research presented in the paper at hand.

\subsection{Related Work}
\label{sec:BGAndRW:RW}
Modeling the software process has been extensively researched over the years. For instance, Bendraou et~al.\ \cite{Bendraou:2010jq} studied six UML-based process modeling languages and evaluated their usability for various use cases. However, over the years, the relevant languages consolidated \cite{kuhrmann2013systematic} resulting in the SPEM \cite{OMG2005} and the V-Modell~XT metamodel \cite{kuhrmann2016use} as the final two actively researched and practically applied large metamodels. However, even though the basically metamodel-free agile methods gained popularity, especially for dependable and large-scale systems, research on software process lines (SPrL) and variability management in such process lines gained more attention \cite{Oliveira:1900cr,simmonds2013variability,de2014software,DBLP:journals/jss/KuhrmannTFRB16,costa2018software} alongside with opportunities to enact and execute software processes to directly support projects based on structured models \cite{Bendraou:2007ja,Oliveira:1900cr,Min:1997cq,kktw2010a,DBLP:journals/scp/KuhrmannKT14}. Costa et~al.\ \cite{costa2018software} provide the most recent state-of-the-art review on the definition of software processes using 26 publications and came to the conclusion that mapping- and rule-based approaches gained most attention. From their research, they derived a need for proper tool support. In \cite{costa2020evaluating}, Costa et~al.\ study the acceptance of such a tool called \emph{Odyssey} in Brazilian software-producing companies. However, Odyssee is, again, a specific, self-contained expert tool that requires users having a specific skillset to use it properly. Notably for complex software process lines, the simple access for end-users is crucial to avoid overheads due to improperly tailored software processes, as it is illustrated by the challenges of adopting process lines in industrial practice \cite{simmonds2013variability}.

The work by Kuhrmann et~al.\ \cite{kktw2010a,DBLP:journals/emisaij/KuhrmannKK13,DBLP:journals/scp/KuhrmannKT14} was specifically focused on defining domain-specific languages to allow for better connecting software process models with project-supporting tools to address the aforementioned issue. Specifically, the PDE and PET\footnote{PDE and PET have been developed in the experimental tools branch of the German V-Modell~XT (cf.\ Table~\ref{tab:CS:ProcessProfile}) and were primarily used to evaluate new features and to explore opportunities for next-generation process support tools.} tool set \cite{kktw2010a} used an \emph{intermediate model} to directly translate metamodel-based software process models into specific target-tool formats, e.g., document templates and project templates for Microsoft Visual Studio and SharePoint. For this, a dedicated meta-metamodel was designed \cite{DBLP:journals/emisaij/KuhrmannKK13}, which allowed for designing process metamodels in an enactable fashion. Source models, like SPEM-/EPF-based processes \cite{OMG2005,eclipse-process-framework} or V-Modell~XT variants \cite{DBLP:journals/jss/KuhrmannTFRB16} could be imported, edited, and exported to various formats, e.g., a Microsoft SharePoint team website including all document templates for a project of a specific kind. However, PDE and PET have been developed using the Visual Studio Language SDKs and followed a modular, yet, ``hard-coded'' approach in which all import and export filters were dedicated modules for expert-tools.

Most work on rich processes that is primarily focused on large-scale projects in regulated environments, such as aerospace, automotive, or public sector, aims at providing more flexibility or even more sophisticated tool support to ease the ``pain'' of using such large-scale models. However, easy access to specifically scoped information required for a specific task in a project was, so far, not in the spotlight. In this paper, we fill this gap by re-thinking complex SPrL-ready process infrastructures to improve the usability and accessibility of task-specific process content by describing a software process ecosystem \emph{as a service}.

\section{An OpenAPI-based Software Process Ecosystem}
\label{sec:SPrLEcosystem}
In this section, we introduce our concept of a software process ecosystem. We start with providing an architecture overview of the original SPrL-infrastructure and the designed target system (Section~\ref{sec:SPrLEcosystem:Architecture}). Based on these architecture considerations, we derive requirements, rules and OpenAPI designs to realize the ecosystem.
\begin{figure*}[!t]
	\centering
	\includegraphics[width=\linewidth]{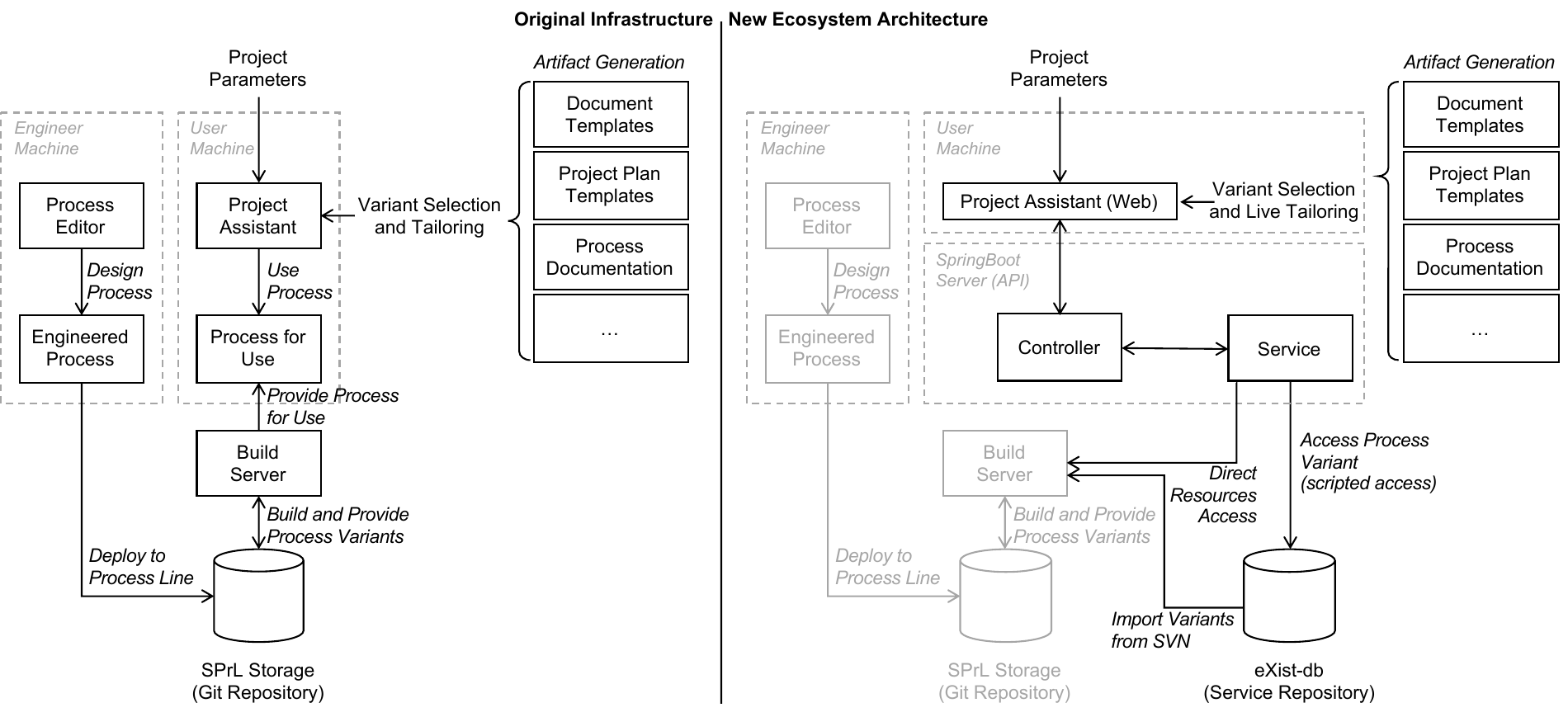}
	\caption{Overall ecosystem architecture (right) compared to the original architecture (left) of the software process line. On the right-hand side, original and unchanged parts are gray to highlight the core components of the new architecture.}
	\label{fig:OverallArchitecture}
\end{figure*}

\subsection{Ecosystem Architecture}
\label{sec:SPrLEcosystem:Architecture}
Figure~\ref{fig:OverallArchitecture} illustrates the overall architecture. In the left-hand part of the figure, the original architecture is shown as a baseline. The SPrL, including all variants, is stored as a set of XML files in a Git repository. This storage server is used by a build server to create the different process variants of the SPrL and to generate the various deployment packages. Starting point for the process variants' development and provision is the editing phase in which a process engineer creates a process variant, pushes it to the storage server that gives it to the build server (following a CI/CD-approach). Process users, e.g., project managers, use a deployment package together with a \emph{project assistant} application. This application is used to conduct the project-specific tailoring on the user's machine, which creates a project-specific process that serves as starting point for exporting various output artifacts, e.g., document templates or a project-specific process documentation.

\paragraph*{New Architecture at a Glance}
The new architecture (Figure~\ref{fig:OverallArchitecture}, right), in its current state, focusses on the provision of process variants to the process users. For this, the process editing activities remain untouched, so does the storage and build server infrastructure. As a first step, the system allows for accessing process variants directly, i.e., an explicit deployment package is no longer required. To support the immediate and easy access, significant modifications in the process tailoring are necessary as the tailoring happens ``live''. In this context, live tailoring means that the project assistant is a web application\footnote{Online (German only): \url{https://padev.dipa.online}} that collects the values for the different tailoring criteria and uses these values as parameters for HTTP requests. A \texttt{Controller} component evaluates the HTTP requests and uses a \texttt{Service} component that selects the required data from an eXist-db that holds pre-processed process variants\footnote{Binary and raw data, e.g., pictures, are still retrieved directly from the SPrL storage.}. The HTTP responses contain the actual process content that can be used for the different exports (generated artifacts).

\paragraph*{Ecosystem User Requirements}
The new architecture as shown in Figure~\ref{fig:OverallArchitecture} emerges from the key requirements listed in Table~\ref{tab:Architecture:Requirements}.
\begin{table}[!t]
\small
	\caption{Overview of the key user requirements to design a web-based software process ecosystem}
	\label{tab:Architecture:Requirements}
	\begin{tabularx}{\linewidth}{lX}
		\toprule 
		\textbf{Req.} & \textbf{Description} \\
		\midrule
		1 & The ecosystem shall provide direct access to a software process. \\
		2 & The ecosystem shall be used as online service (no local software). \\
		3 & The ecosystem shall abstract from the technical process model. \\
		4 & The ecosystem shall provide a future-proof platform\newline (reduction of technical dependencies). \\
		\bottomrule 
	\end{tabularx}
\end{table}
As shown in Table~\ref{tab:Architecture:Requirements}, one of the key objectives of the process ecosystem is to provide a software process ``on demand'' without requiring process users to install comprehensive software packages, e.g., for project assistants and exporters. The goal is to provide the software process as a service in terms of providing access to a process repository from which users select a process of their liking, configuring the selected process according to their context-specific needs, and exporting required process documentation or document templates based on their configuration. As the ecosystem still relies on the original SPrL backbone, users have the opportunity to either select a specific (``old'') process version or just going for the most recent process. Process providers have the opportunity to easily update a process or to add new features to process-supporting tools.

These user requirements illustrate the need to have an easy and quick access to a software process if needed, that is, if a new project starts, project managers are not required to install a complex client system ``just'' to get a few document templates. Instead, users shall characterize their project and shall be enabled to browse their tailored process online to obtain the required process artifacts. Hence, the presented ecosystem allows for a small client-side footprint, a faster access to the required process parts, and a reduced learning curve (no specific local process tools).

\subsection{Modeling Approach}
\label{sec:SPrLEcosystem:ModelingApproach}
In this section, we present the modeling approach, which (i) demonstrates how a process (meta-)model can be made accessible online, (ii) documents the constraints to be considered in the modeling approach, and (iii) illustrates the meta-model transition into a technical infrastructure. 

\subsubsection{Preparatory Considerations}
\label{sec:SPrLEcosystem:ModelingApproach:Preparation}
Complex software process models are usually grounded in a metamodel, which, itself, can become fairly complex, e.g., the SPEM metamodel \cite{OMG2005}. For this, to bring a software process online, a respective ecosystem requires an appropriate process language to translate a given process into a generic API, which can be accessed via an online service.
\begin{table}[!t]
\small
	\caption{Overview of key steps in the process language design}
	\label{tab:Architecture:ProcessLanguageDesign}
	\begin{tabularx}{\linewidth}{lX}
		\toprule 
		\textbf{Step} & \textbf{Description} \\
		\midrule
		1 & Development of a conceptual metamodel of the relevant software process model(s) from a metamodel or a number 
			of exemplars. \\
		2 & Identification of the relevant process model entities (process elements) in the conceptual metamodel. \\
		3 & Identification of the relevant (navigable) associations among the process elements in the conceptual metamodel. \\
		4 & Design of the API to access the process content (selection of process elements and navigation 
			between the process elements).\\
		\bottomrule 
	\end{tabularx}
\end{table}

Table~\ref{tab:Architecture:ProcessLanguageDesign} summarizes the key steps in the process language development. In the first step, a \emph{conceptual metamodel} is developed, either from an existing (technical) metamodel or extracted from a number of process exemplars. This conceptual model serves two major purposes: (i) the relevant process elements have to be identified to answer the question for the browsable entities of the process model. In order to browse the process model, (ii) associations among the different entities have to be identified and modeled. In the course of identifying those entities and associations, modeling conventions need to be defined. Modeling conventions are later used to develop a proper mapping of the conceptual metamodel to a particular technical infrastructure. For instance, if a \emph{WorkProduct} in the original process model is defined as a \emph{composite process element}, the container element as well as the composed process elements have to be identified as entities, and connections have to be established, such that the container knows its contained parts and the parts know their container. For this, \emph{association classes} with respective navigation properties have to be defined. Since we opted for UML as conceptual modeling language, given the example above, we need to define that a \emph{UML composition association} needs to be semantically established between the parts and the container and the navigation properties have to be set accordingly. Eventually, this approach leads to a limitation of the process language elements that, finally, are used to define the web-based service API.

\subsubsection{Process Language Design Rules}
\label{sec:SPrLEcosystem:ModelingApproach:DesignRules}
In the following, we present the process language design rules, which we illustrate with selected examples. The design rules initially refer to the design of the conceptual metamodel and, therefore, refer to the UML modeling concepts. These design rules lay the foundation for the API development of the \texttt{Controller} and \texttt{Service} components (Figure~\ref{fig:OverallArchitecture}).

\paragraph{UML Modeling Concept Selection}
The primary concept used for the metamodel development is the \emph{UML Class Diagram}. Yet, not all notation elements for class diagrams are necessary and, therefore, the following notation elements are selected to develop the conceptual metamodel: for the \emph{type/class level}, \texttt{public}, \texttt{protected}, and \texttt{private} fields, and type generalizations (inheritance) are allowed for use. For associations, the association types \texttt{composition}, \texttt{aggregation}, and (directed) association are allowed for use. In subsequent sections, we describe the language design rules for the selected modeling concepts.

\begin{figure}[!t]
	\centering
	\includegraphics[width=\linewidth]{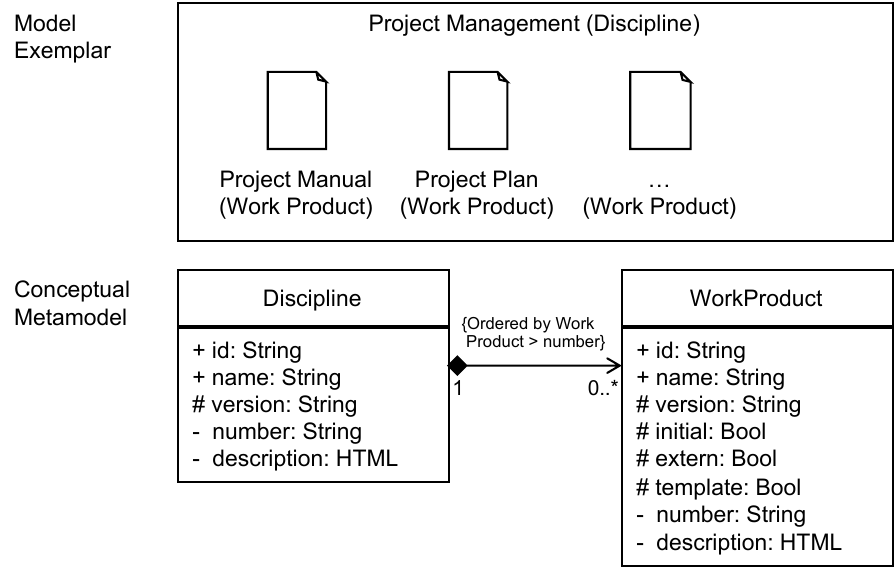}
	\caption{Providing the set of work products associated with a discipline}
	\label{fig:ProduktZuDisziplinExample}
\end{figure}

\paragraph{Association Navigation Rules}
Navigation rules refer to a set of rules applied to associations between types/classes to make the metamodel instances (the process models) browsable, i.e., starting with a given process element $A$, the question is answered \emph{``what belongs to $A$?''}. For this, identified process elements (Table~\ref{tab:Architecture:ProcessLanguageDesign}, step 2) become \emph{Endpoints} that can be addressed and thus reached by the \texttt{Service} component (Figure~\ref{fig:OverallArchitecture}). Navigability now refers to the navigation between endpoints and is expressed via the \texttt{composition} association type in the conceptual metamodel (Figure~\ref{fig:ProduktZuDisziplinExample}). The other two association types are only used for creating hyperlinks for in-model navigation, but, do not become browsable endpoints.

\paragraph{Type/Class Rules}
Type-/class-level rules refer to the interpretation of model entities (process elements) in the type system design, specifically, types that can be retrieved via an API as endpoints. The process elements represented by endpoints have attributes (Figure~\ref{fig:ProduktZuDisziplinExample}), and some endpoints represent process elements including composed sub-elements. Yet, not all attributes of the aggregated process elements are relevant and, therefore, must not be \emph{visible} at an endpoint. The handling of attributes via their visibility is summarized in Table~\ref{tab:Architecture:ProcessLanguageDesign:EndpointAttributeHandling}.
\begin{table}[!t]
\small
	\caption{Handling of process element attributes visibility}
	\label{tab:Architecture:ProcessLanguageDesign:EndpointAttributeHandling}
	\begin{tabularx}{\linewidth}{lX}
		\toprule 
		\textbf{Visibility} & \textbf{Description} \\
		\midrule
		\texttt{public}    & Public attributes are always returned by the API endpoint. In case the selected process element
							 refer further process elements, those elements are returned as sub-elements including their
							 respective public attributes. \\
		\texttt{protected} & Protected attributes are returned from selected process elements and aggregated sub-elements
							 only (\texttt{aggregation} type associations). \\
		\texttt{private}   & Private attributes are only available if a specific process element is directly accessed by
							 its \texttt{id} (single-element access via endpoint). \\
		\bottomrule 
	\end{tabularx}
\end{table}

\paragraph{Example for Resolving Composite Model Structures}
To illustrate the above rules, we use Figure~\ref{fig:ProduktZuDisziplinExample} as an example. The model exemplar (upper part of the figure) leads to the conceptual metamodel fragment (lower part), which consists of two process elements \texttt{Discipline} and \texttt{WorkProduct} with respective attributes. These two process elements become browsable endpoints, and both process elements are linked through a \texttt{composition} association type. That is, the system (Figure~\ref{fig:OverallArchitecture}) allows for retrieving all work products for a given discipline. For illustration purposes, we utilize the final OpenAPI components just here to show how the concepts are finally realized.
As a first step, we design the process elements. Listing~\ref{lst:WorkProductsPerDiscipline} shows the technical realization of the \texttt{Discipline} class from the conceptual metamodel. The class' attributes are modeled as XML attributes. The \emph{composition} association between the \texttt{Discipline} the a \texttt{WorkProduct} is realized as an embedded XML element, which lists further XML nodes per linked work product. 
\begin{lstlisting}[language=XML,basicstyle=\footnotesize\ttfamily,caption={Endpoint model of the work products associated with a discipline},label=lst:WorkProductsPerDiscipline]
<?xml version="1.0" encoding="UTF-8"?>
<response>
  <Discipline id="string" version="string" name="string">
    <WorkProducts>
      <WorkProduct id="string" name="string"></WorkProduct>
    </WorkProducts>
  </Disciple>
</response>
\end{lstlisting}

This XML message is retrieved when (technically) calling the \texttt{Service} component (Figure~\ref{fig:OverallArchitecture}) as follows:
\begin{lstlisting}[language=XML,basicstyle=\footnotesize\ttfamily]
api/discipline
\end{lstlisting}
According to the rules defined in Table~\ref{tab:Architecture:ProcessLanguageDesign:EndpointAttributeHandling}, only public type attributes are returned to the caller. If the full data package is necessary, a specific element including all attributes can be retrieved by calling:
\begin{lstlisting}[language=XML,basicstyle=\footnotesize\ttfamily]
api/discipline/{disciplineId}
\end{lstlisting}
This call returns the full data element, which is shown in Listing~\ref{lst:Discipline}. Still, the \texttt{WorkProduct} element only contains selected public attributes resulting from the application of the rules shown in Table~\ref{tab:Architecture:ProcessLanguageDesign:EndpointAttributeHandling}. The \texttt{WorkProduct} element is, therefore, the \emph{Endpoint} for the work product type and, if the detailed data package for a work product is required, the API call must include the respective work product's \texttt{Id}.
\begin{lstlisting}[language=XML,basicstyle=\footnotesize\ttfamily,caption={Endpoint model of a discipline represented as XML message},label=lst:Discipline]
<?xml version="1.0" encoding="UTF-8"?>
<Discipline id="string" version="string" name="string">
  <Number>0</Number>
  <Description>Description text as HTML...</Description>
  <WorkProducts>
    <WorkProduct id="string" name="string"></WorkProduct>
  </WorkProducts>
</Discipline>
\end{lstlisting}

Eventually, the metamodel, and the resulting data- and endpoint models are used to define the OpenAPI interface (Section~\ref{sec:SPrLEcosystem:ModelingApproach:APIDesign}). The API call for the scenario from Figure~\ref{fig:ProduktZuDisziplinExample} would be:
\begin{lstlisting}[language=XML,basicstyle=\footnotesize\ttfamily]
api/discipline/{disciplineId}/workproduct
\end{lstlisting}
Calling the \texttt{Service} component (Figure~\ref{fig:OverallArchitecture}) this way, the system returns a list of all work products associated with a discipline in a response message (Listing~\ref{lst:WorkProductsPerDiscipline}).

\paragraph{Example for Resolving Named Associations}
While the first example illustrated how complex, composite model structures are dynamically resolved, in this section, we show the handling of ``ordinary'', yet named, association types. 

\begin{figure}[!t]
	\centering
	\includegraphics[width=\linewidth]{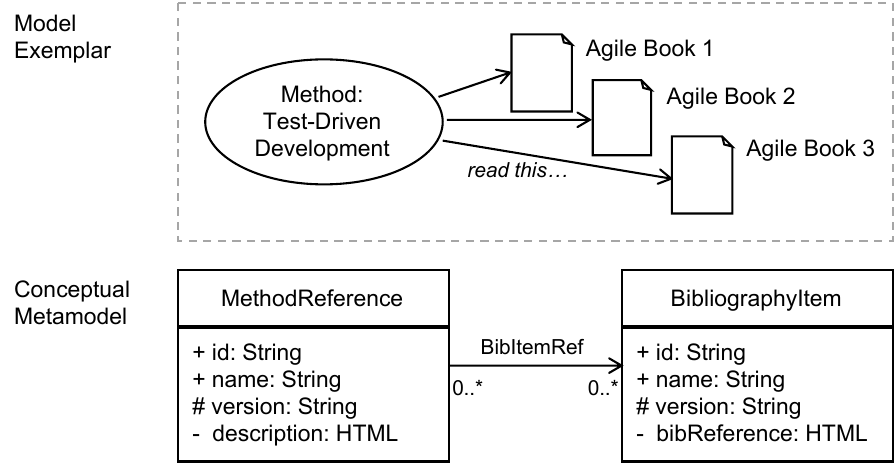}
	\caption  {Providing bibliography items to supplement a method/practice option}
	\label{fig:MethodRef}
\end{figure}

Figure~\ref{fig:MethodRef} shows the example of an optional method practice to be applied to selected tasks in a software development project. For this method/practice, the process model could state: \emph{``If you selected an agile software development approach, consider using test-driven development (TDD).''} To help process users understand what it means implementing TDD in their project-specific process, the process model can provide a so-called \emph{method reference} in which such optional methods and practices are named---usually including a short description. If further information is required, complementing literature is referred. 

The resulting conceptual metamodel for this example is shown in Figure~\ref{fig:MethodRef} (lower part). A type \texttt{MethodReference} including its attributes is defined, which refers to a bibliography of which the items are modeled as \texttt{BibliographyItem}. A named association \texttt{BibItemRef} links a method reference to a bibliography item. Both types are \emph{endpoints}, and if a process user wants to get all bibliography items associated with a method reference, the endpoint has to be called at the \texttt{Service} component. Like in the first example, to get all bibliography items from the model, the \texttt{Service} component needs to be called as follows:
\begin{lstlisting}[language=XML,basicstyle=\footnotesize\ttfamily]
api/methodreference/{methodreferenceId}/bibitemref
\end{lstlisting}
Associated model elements are resolved via the named association, whereas the association name defines the \emph{return type} of the query. The return type then holds the public attributes of the referred type (see Table~\ref{tab:Architecture:ProcessLanguageDesign:EndpointAttributeHandling}). In case the association end has a multiplicity of 0..* or 1..*, an extra container XML element is generated to provide a list of XML elements as shown in Listing~\ref{lst:MethodRef}
\begin{lstlisting}[language=XML,float,floatplacement=t,basicstyle=\footnotesize\ttfamily,caption={XML response message for querying bibliography items for a given method reference},label=lst:MethodRef]
<?xml version="1.0" encoding="UTF-8"?>
<response>
  <MethodReference id="string" version="string" 
                               name="string">
    <BibItemRefs>
      <BibItemRef id="string" name="string"></BibItemRef>
    </BibItemRefs>
  </MethodReference>
</response>
\end{lstlisting}

This example, together with the first example, shows the methodological approach in the modeling approach: all process elements of interest are translated into types, which can serve as endpoint for the ecosystem's \texttt{Service} component. Dependencies between these elements are modeled as associations of specific types, which are used to connect the endpoints and, eventually, generate XML messages at the technical level. Thus, particularities of an \emph{arbitrary} process metamodel can be abstracted and locked away behind a generic web-based API. This paves the way for mapping different process metamodels to the ecosystem without bothering process users with specific technical aspects of the respective processes.

\paragraph{Custom Rules}
The navigation rules introduced above build the foundation for defining \emph{custom rules}, which can be defined reusing the basic rules. One example for such a custom rule is the modeling approach for the \emph{process tailoring}. Process tailoring is at the metamodel-level defined by a \texttt{ProjectCharacteristic} type, which is connected to a set of \texttt{Value} objects that represent specific values for a characteristic, e.g., \emph{Project Size = Large} with \emph{Project Size} being an exemplar of \texttt{ProjectCharacteristic} and \emph{Large} being an exemplar of \texttt{Value}. Each value is associated with a number of process elements (types), e.g., process modules, workflows, and so forth, that are linked to a specific project profile characterized by a set of values. Hence, the tailoring can be implemented by providing API calls with a map of parameters, which act as a real-time filter to access certain process elements. This design approach has significant implications as the process tailoring is executed in real-time on a per model-access call. That is, a project-specific tailoring is not executed ``en block'' once in the beginning of a project. Rather, based on a saved profile, every model-access call carries the full parameter set and, thus, the tailoring is executed on the model during the computation of the response message.

\subsubsection{API Design}
\label{sec:SPrLEcosystem:ModelingApproach:APIDesign}
Having the conceptual and technical metamodel designs available, the definition of the endpoints in the OpenAPI is straightforward. Listing~\ref{lst:ProduktDisziplinExampleYAML} shows the API-definition for the endpoint to retrieve a \texttt{Discipline} message as shown in Listing~\ref{lst:Discipline}. The list of work products belonging to a discipline (Listing~\ref{lst:Discipline}) is generated by referring to the type \texttt{WorkProduct} for which the attributes of interest are defined in a list of \texttt{ref}-elements.

\subsubsection{Technical Realization}
The technical realization of the API's backbone was done using the \emph{Spring Boot} infrastructure\footnote{Spring Boot online: \url{https://spring.io/projects/spring-boot}}, which is a Java-based REST API that communicates via the HTTP protocol. Hence, the ecosystem's backbone is fully hidden behind a web-based service interface thus removing any realization-specific dependencies. On the server-side, the \texttt{Service} and \texttt{Controller} components (Figure~\ref{sec:SPrLEcosystem:Architecture}) are implemented using Java. The systems data storage is the XML database eXist-db\footnote{eXist Solutions online: \url{http://exist-db.org/exist/apps/homepage/index.html}}.

\begin{lstlisting}[language=XML,float,floatplacement=t,basicstyle=\footnotesize\ttfamily,caption={OpenAPI specification to access the endpoint defined in Listing~\ref{lst:Discipline}},label=lst:ProduktDisziplinExampleYAML]
Discipline:
 type: object
 xml:
  name: 'Discipline'
 required:
  - id
  - version
  - name
 properties:
  id:
   type: string
   xml:
    attribute: true
  version:
   type: string
   xml:
    attribute: true
  name:
   type: string
   xml:
    attribute: true
  number:
   type: integer
  description:
   $ref: './typeHtml.yaml#/typeHtml'
  WorkProduct:
   type: array
   items:
    type: object
    xml:
     name: 'WorkProduct'
    properties:
     id:
      $ref: './[file].yaml#/WorkProduct/properties/id'
     name:
      $ref: './[file].yaml#/WorkProduct/properties/name'
   xml:
    wrapped: true
    name: 'WorkProducts'
\end{lstlisting}

\section{Case Study}
\label{sec:CaseStudy}
In this section, we provide a proof-of-concept validation. For this, we adopt the \emph{case study} instrument as described in Wohlin et~al.~\cite{Wohlin-C.:2012uqp} to (i) illustrate the context and implementation and (ii) to evaluate the reasonability of the presented concept. In subsequent sections, we present the objective and the research questions (Section~\ref{sec:CaseStudy:ObjAndRQs}), the case selection and description (Section~\ref{sec:CaseStudy:Case}), the data collection and analysis (Section~\ref{sec:CaseStudy:DataCollection}), the presentation of the results (Section~\ref{sec:CaseStudy:Results}), and, finally, a discussion of the results and the validity constraints in Section~\ref{sec:CaseStudy:DiscussionAndTTV}.

\subsection{Objective and Research Question}
\label{sec:CaseStudy:ObjAndRQs}
Our overall goal is \emph{to provide comprehensive software process ecosystems including models and tools as an easy-to-use Web-based service}. To evaluate the developed concept, we pose the following research questions:
\begin{itemize}
	\item[\textbf{RQ1:}] \emph{To what extent can a software process ecosystem be realized as a Web-based service?} This research question aims to study the conceptual and technical opportunities of providing software process ecosystems as a web-based service by analyzing the capabilities of a given software process and comparing these to the realization described in the paper at hand. We aim to provide an analysis of the individual parts of the process ecosystem and their realizability in a web-based service environment.

	\item[\textbf{RQ2:}] \emph{Does the Web-based service provide benefits compared to traditional process model use?} This research question aims to evaluate whether the concept presented in the paper at hand provides benefits. For this, we conduct an expert interview with tool vendor and  process user representatives---all experienced in the original process that was used for the transition to the web platform. We aim to analyze advantages and/or disadvantages of the concepts and implementations presented in this paper.

\end{itemize}

\subsection{Case Selection and Description}
\label{sec:CaseStudy:Case}
As case, we opt for the \emph{V-Modell~XT}, which is the IT development standard in German public authorities\footnote{V-Modell XT reference model: \url{https://www.cio.bund.de/Webs/CIO/DE/digitaler-wandel/Achitekturen_und_Standards/V_modell_xt/v_modell_xt-node.html}}. The V-Modell~XT is a synonym for a comprehensive and large-scale software process line \cite{DBLP:journals/jss/KuhrmannTFRB16} that consists of a base model, the so-called \emph{reference model}, and a number of organization-specific adaptations, the so-called \emph{variants}. Furthermore, the process line includes several tools, namely an editor component, a project assistant supporting project managers in setting up and tailoring their project-specific processes, and a comprehensive CI/CD-based build environment to manage and deploy the different process variants. To realize the proof-of-concept, we use the reference process \emph{V-Modell~XT} in the versions 2.3 and 2.4 together with the \emph{V-Modell~XT Bund} as a derived variant within the process line\footnote{Please note that to start the realization activities, the full conceptual and technical metamodels need to be in place. Due to the complexity of the metamodel and the space limitations, we cannot present a full picture of metamodel in the paper at hand and, therefore, refer to the fragments shown in Figure~\ref{fig:ProduktZuDisziplinExample} and Figure~\ref{fig:MethodRef} as examples.}.

Being a practically used software process, there are tool vendors and process users available for evaluation purposes. Specifically, one person from the tool-vending company was recruited for an interview on the technical parts and, respectively, two persons from the process-owning organization were selected as interviewees for the process-use parts.

\subsection{Data Collection and Analysis}
\label{sec:CaseStudy:DataCollection}
The data collection procedure consists of two parts: (i) to study RQ1, we develop a conceptual and technical profile of the process ecosystem under consideration, which we apply to the new concepts and analyze to which extend the new approach covers the original profile. (ii) We create two interview guidelines to get insights from the experts to answer RQ2.

\subsubsection{Conceptual and Technical Process Profile}
\label{sec:CaseStudy:DataCollection:Part1}
Table~\ref{tab:CS:ProcessProfile} provides an overview of the conceptual and technical profile of the selected software process. This profile is used to evaluate the design and realization of the web-based service presented in the paper at hand. For each of the evaluation elements listed in Table~\ref{tab:CS:ProcessProfile}, we analyze the design and realization and, if fulfilled, we name the actual realization. From the number of matches, we conclude the reasonability of the conceptual and technical realization of a software process ecosystem.
\begin{table}[!t]
\small
	\caption{Conceptual and technical profile of the V-Modell XT}
	\label{tab:CS:ProcessProfile}
	\begin{tabularx}{\linewidth}{lX}
		\toprule 
		\textbf{Element} & \textbf{Description} \\
		\midrule
		Metamodel & Yes (XSD) \\
		Machine-readable Model & Yes (XML, HTML) \\
		Editor Component & Yes (V-Modell XT Editor) \\
		Tailoring Component & Yes (V-Modell XT Project Assistant) \\
		Process-line Management & Yes (Editor, Build Environment) \\
		Document Generation & -- Document Templates\newline
							  -- Process Documentation \newline
							  -- Project Plan Template \newline
							  -- Compliance Reports (exp.)*\newline
							  -- Project Templates (exp.)* \\
		\bottomrule 
	\end{tabularx}
	\footnotesize\flushleft
	*experimental tools in complementing activities. The compliance reports are used to analyze the process line as a whole \cite{DBLP:journals/jss/KuhrmannTFRB16}, the project templates have been created as opportunity analysis for Microsoft SharePoint and the Microsoft Team Foundation Server \cite{DBLP:journals/emisaij/KuhrmannKK13}.
\end{table}

\subsubsection{Practitioner Survey}
\label{sec:CaseStudy:DataCollection:Part2}
To get insights regarding the feasibility of our concept's implementation, we interviewed experts in the process-supporting tools' development and experienced process owners/users. The experts were selected as follows: for the tool-development perspective, we interviewed the team leader of the development team, who was responsible for the original tool infrastructure (Figure~\ref{fig:OverallArchitecture}) and, therefore, was able to evaluate the new approach to support the process in comparison with the original tool infrastructure. For the process-user perspective, we interviewed two experts---one being a process owner of an organization-specific process variant, and both experts are experienced process users, in terms of project management and consulting.

\begin{table*}[!ht]
\small
	\caption{Interview/survey guideline used for the expert interview (all experts have been asked the same questions, only the questions 2 and 3 were adjusted to reflect the interviewees' roles)}
	\label{tab:CS:PractSurvey}
	\begin{tabularx}{\linewidth}{llllX}
		\toprule 
		\textbf{Category} & \textbf{No.} & \textbf{Scale*} & \textbf{Questions (tool)} & \textbf{Questions (user)} \\
		\midrule
		Use Case		& 1 & FT & What is the use case/goal of the tool? & \\
		Feature Set	& 2 & MC & Which features are implemented in the new tool? &
						     Which features do you use in the new tool?\\
					& 3 &  LI & Please evaluate the ease of implementation of the tool.  &
					           Please evaluate the ease of use of the tool. \\
		Evaluation	& 4 & FT & Please provide a brief explanation of the advantages of the new tool. & \\
					& 5 & FT & Please provide a brief explanation of the disadvantages of the new tool. & \\
					& 6 & FT & Is there anything else you want to share with us? & \\
		\bottomrule 
	\end{tabularx}
	\footnotesize\flushleft
	*FT: free-text answer, MC: multiple-choice answer, LI: Likert-scale (very easy--very hard; na)
\end{table*}

The interview/survey guideline is shown in Table~\ref{tab:CS:PractSurvey}. The table shows that all experts have been asked the same questions, only with minor adjustments to address their specific roles. Specifically, all experts were asked to briefly describe the use case of the newly developed tool, which provides access to the ecosystem to work out, if the original use case was still properly addressed or if a new use case was targeted. Since the original tool provided a specific set of features, we interviewed the tool developer if all the features are still available and how easy it was to implement those features. Furthermore, we asked a follow-up question for the ease of implementation in comparison to the original tool. Regarding the process users, we asked the same questions, yet from the perspective of the features' use. Finally, all experts were asked to provide a short statement regarding the advantages and disadvantages of the new tool, if possible in comparison to the original tool and infrastructure.

The guideline was compiled by one researcher, quality-assured by a second researcher, who conducted the interviews. The interview with the tool developer was conducted in-person in a video conference, the interview with the process users was implemented as a survey after the process users have been instructed regarding the questionnaire. The interview protocols have been transcribed (tool developer) and extracted into a spreadsheet (process users). A researcher, who was not involved in the interviews analyzed the answers, and a second researcher conducted the quality assurance of the analyzed data.

\subsection{Results}
\label{sec:CaseStudy:Results}
In this section, we present the results of our proof-of-concept evaluation. In Section~\ref{sec:CaseStudy:Results:RQ1}, we provide the analysis of the completion of the conceptual and technical profiles and Section~\ref{sec:CaseStudy:Results:RQ2} provides the analysis and interpretation of the practitioner survey.

\subsubsection{Evaluation of the Conceptual and Technical Profile}
\label{sec:CaseStudy:Results:RQ1}
Table~\ref{tab:CS:ProcessProfile:Evaluation} provides an overview of the evaluation of the conceptual and technical profiles derived from the original process model infrastructure to analyze which parts of the profile are realized and evaluated in practice. Specifically, the metamodel, the machine-readable process model, and the tailoring component are fully transitioned to the new ecosystem. As outlined in Section~\ref{sec:SPrLEcosystem:ModelingApproach:DesignRules}, the original metamodel was reengineered such that the purely XML-schema-based metamodel was translated into a conceptual metamodel, which was used to derive a set of realization/implementation rules steering the data storage. Other than in the original process infrastructure, the data storage is now based on a database system (instead of a file system). As the process-line management (including the complex build infrastructure) was not touched in this development stage, the data im-/export was realized such that the ecosystem's database imports their data from that build environment and, thus, is seamlessly integrated with the process-line backend, which also includes the process authoring parts (editor component). 
\begin{table*}[!ht]
\small
	\caption{Evaluation of the conceptual and technical profile of the V-Modell XT}
	\label{tab:CS:ProcessProfile:Evaluation}
	\begin{tabularx}{\linewidth}{lp{40mm}lX}
		\toprule 
		\multicolumn{2}{l}{\textbf{Profile (Original Infrastructure)}} & \multicolumn{2}{l}{\textbf{Profile (Evaluation new Infrastructure)}} \\
		\textbf{Element} & \textbf{Description} & \textbf{Eval.} & \textbf{Explanation/Rationale} \\
		\midrule
		Metamodel & Yes (XSD) 							& Yes & Full support: the metamodel was reengineered into a 
				conceptual metamodel, a technical/implementation-related metamodel, and an API to support the implementation
				of the infrastructure. The respective concepts can be taken from Section~\ref{sec:SPrLEcosystem}.\\
		Machine-readable Model & Yes (XML, HTML) 			& Yes & The process model is realized as XML data structure 
				stored in a database. The non-XML data, such as binary files (figures, document templates) are separately 
				stored and made accessible through the ecosystem's service components. The respective implementation rules
				are described in Section~\ref{sec:SPrLEcosystem:ModelingApproach:DesignRules} and 
				Section~\ref{sec:SPrLEcosystem:ModelingApproach:APIDesign}. \\
		Editor Component & Yes (V-Modell XT Editor) 			& No  & Not relevant in this development stage. \\
		Tailoring Component & Yes (V-Modell XT Project Assistant) & Yes & A new tool was developed with respect to the
				particularities of the new ecosystem based on the requirements listed in Table~\ref{tab:Architecture:Requirements}. 
				The new tool's evaluation can be taken from Section~\ref{sec:CaseStudy:Results:RQ2}. \\
		Process-line Management & Yes (Editor, Build Environment) & No & According to the infrastructure, the original build
				environment is still used in this development stage (Figure~\ref{fig:OverallArchitecture}).\\
		Document Generation & -- Document Templates\newline
							  -- Process Documentation \newline
							  -- Project Plan Template \newline
							  -- (experimental tools) 	& Partially & The new tool provides a certain feature set, which
				is evaluated in Section~\ref{sec:CaseStudy:Results:RQ2}. Further experimental tools have not been considered
				in the current development stage.\\
		\bottomrule 
	\end{tabularx}
\end{table*}

End-user access to the ecosystem is provided through the assistant component, which replaces the original application and is, itself, a web-based application instead of a desktop-based system. In order to provide the process users with a satisfying set of features, the goal of the development of this application was to provide at least those features that end-users are used to. The evaluation of the feature sets can be taken from Section~\ref{sec:CaseStudy:Results:RQ2}. This includes, among other things, different export features, such as the generation of document templates based on a process model tailored to a project-specific setting. Experimental tools, however, have been excluded from our work and are, therefore, not subject to the work presented in the paper at hand.

\vspace{0.35em}
\noindent
\fbox{
	\begin{minipage}{0.955\columnwidth}
		\small
		\textbf{Summary:} The evaluation of the conceptual and technical profiles showed that the developed concepts can be translated into web-based ecosystem, i.e., we demonstrated the realizability of a complex, metamodel-based process framework including the complementing tool infrastructure as a web-based service. Furthermore, due to the selection of a specific use case for the transition into the new ecosystem, we could also show a migration path that, eventually, will be able to cover a step-wise migration of the entire legacy-system.
	\end{minipage}
}

\subsubsection{Practitioner Survey}
\label{sec:CaseStudy:Results:RQ2}
To answer the second research question, we conducted interviews with three experts based on the guideline shown in Table~\ref{tab:CS:PractSurvey}. We present the findings based on the experts' statements alongside with our interpretation.

\paragraph{Tool-vendor Perspective}
We recruited the team leader for the development of the original and new tool set, since this person was expected to be able providing the most reliable statements regarding the actual development effort, benefits, and issues. Regarding the first question, the interviewee stated that \emph{``[\ldots] the overall use case has not been changed. The primary goal was to make the general use of the tool easier and to avoid tool installation effort when updating the process framework [\ldots]''}. Regarding the implemented features, the difficulty of implementing the features using the new ecosystem-backend, the interviewee stated \emph{``'[\ldots] basically all features are available in the new version and the development effort is comparable [\ldots]'}. The simple integration of new versions and updated features was mentioned a significant improvement compared to the original version of the tool, yet, as a disadvantage, the complex query construction to access the ecosystem's \texttt{Service} component was mentioned: \emph{``[\ldots] it was quite complex before, and a more convenient solution would have been appreciated [\ldots]''}.

\vspace{0.35em}
\noindent
\fbox{
	\begin{minipage}{0.955\columnwidth}
		\small
		\textbf{Interpretation:} The development of process-supporting software tools is quite complex anyways, since those tools are expert tools. The interview with the tool vendor, however, showed that (i) all feature could be made available, (ii) the difficulty of the implementation is comparable with the development effort required before, and (iii) the flexibility of the tool as been increased while, at the same time, platform dependencies have been reduced.
	\end{minipage}
}

\paragraph{End-user Perspective}
We recruited two experts that are experienced in the process framework (V-Modell~XT), and that are also experienced in the use of the infrastructure including the respective tools. Both experts agree on an unchanged use case, i.e., provision of process users with project-specific processes to organize and steer their projects. Both experts use the features: \emph{conducting the process tailoring}, \emph{generation of project-specific process documentation}, and \emph{export of work product templates from a tailored process model}. One of the experts, who's also acting as a process owner, additionally, uses the features \emph{selection of a process variant}, \emph{selection of a process version}, and \emph{utilization of the tool for training purposes}. However, neither of the interviewed experts used the \emph{project plan template export} feature other than for testing purposes. Both experts, for the features they used, agree that the usability of five out of the six of the features in the new tool is very good (very easy to use), while one feature (\emph{project plan template export}) has shown no improvement compared to the original version. On the usability, the first expert states: \emph{``[\ldots] the most obvious simplification is the `just use' approach and that no further installation including all the hurdles is necessary anymore [\ldots]''}. Also, both experts agree that the usability of the web-based tool is much higher than before, only the local storage approach for exported document templates was considered subject to future improvement.

\vspace{0.35em}
\noindent
\fbox{
	\begin{minipage}{0.955\columnwidth}
		\small
		\textbf{Interpretation:} The two end-user experts appreciated the change in the usage approach, i.e., using the process models as web-based service. From the perspective of the feature use, both experts agree that the accessibility of the process has been improved, since the requirement to run through a complex desktop installation procedure with numerous supporting components does not exist anymore. Both experts agree that the web-based design and delivery approach constitutes a significant improvement from the user's perspective. That is, the re-design of the process framework including the supporting tools towards a web-based system was appreciated.
	\end{minipage}
}

\subsection{Discussion and Threats to Validity}
\label{sec:CaseStudy:DiscussionAndTTV}
In this section, we answer the research questions and discuss our finding in comparison to the initial requirements (Table~\ref{tab:Architecture:Requirements}), before we discuss the threats to validity.

\subsubsection{Answering the Research Questions}
The first research question was concerned with the extent to which a complex process ecosystem can be transitioned into a web-based service. The evaluation of the conceptual and technical profiles (Section~\ref{sec:CaseStudy:Results:RQ1}) showed a full implementation of all features selected for this development stage. Just this stage already reached a high coverage of the overall conceptual and technical profile created for the original process ecosystem (Table~\ref{tab:CS:ProcessProfile:Evaluation}). Since the selected part of the system was fully realized and the effort required for the implementation was not higher compared to the original system, our concepts presented in Section~\ref{sec:SPrLEcosystem:ModelingApproach} lay the foundation for further transition steps.

The second research question was concerned with the benefits the new ecosystem is supposed to provide to the process users. The three expert interviews showed that the benefits are (i) the simple, just-in-time use through the web-application (Section~\ref{sec:CaseStudy:Results:RQ2}) and (ii) the easy access to process content and functionality without the need of running through complex installation procedures for process-supporting expert tools. Hence, we consider the new ecosystem beneficial for process users, since they are not required to mingle around with complex tools, just for exporting a few document templates. This finding, in-line with the finding regarding RQ1, motivates further steps towards a full transition, e.g., by transitioning the process authoring tools to the new ecosystem.

\subsubsection{Discussion of the Findings}

In Table~\ref{tab:Architecture:Requirements}, we listed the key requirements motivating our work. The evaluation with the experts shown above, shows those four key requirements fulfilled. From the developers' perspective, ``just'' the backend changed with a new web-based API, but it was stated that the general development effort was kept at the same level. That is, implementing further process-supporting tools with an acceptable effort is possible. Since the API can be retrieved via the OpenAPI infrastructure, the general effort for providing new (experimental) tools could be fostered as no large backend infrastructure needs to be provided to developers---the web-based ecosystem becomes a ``simple'' service. 

From the end-users' perspective, the new ecosystem including the new tools constitutes an improvement in terms of usability of the assistance tools (small footprint, no further installation routines) and the accessibly of the process (simple access via the web). For this, we consider the requirements (Table~\ref{tab:Architecture:Requirements}) met and, furthermore, we argue that the new ecosystem is ready for a transition of further process framework components, such as the process authoring tools. However, it must be admitted that the current work is primarily driven by process-user use cases, and primarily aimed at improving the accessibility and usability of the final process for utilization in software projects. The reduction of complexity at the end-user's side will likely raise the ``price'' at the process engineering side for which a deeper elaboration is subject to future work.

\subsubsection{Threats to Validity}
The proof-of-concept validation presented before is evaluated regarding the threats to validity according to Wohlin et~al.~\cite{Wohlin-C.:2012uqp}. The \emph{internal validity} is threatened by the small size of the expert group and their recruiting procedure. To evaluate our concepts and implementations, a specific kind of expert is necessary and such experts are rare. To improve the validity, we opted for separating the data collection and analysis procedures: the interview guideline design and the data analysis was done by a researcher not involved in the actual expert interaction, while the team members involved in the development project only collected, yet not analyzed or interpreted the interview data.
The \emph{conclusion validity} might be affected by the case selection. Since the process infrastructure selected for the case is large with an almost 20-year history, there is the risk that a part of the infrastructure was chosen, which was too easy to handle and, thus, is not representative for the entirety of the infrastructure. Our conclusions could thus be overly positive.
The \emph{external validity} is affected by the specificity of the objectives and the case. The management and the provision of tool-support for comprehensive software process lines are specific expert activities. Even though previous research suggests that intermediate models successfully work for complex transition and conversion tasks, still, the generalizability of our results can only be improved by extending the number of cases studied.

\section{Conclusion}
\label{sec:Conclusion}
In this paper, we presented our approach to re-design a complex SPrL infrastructure into a web-service-based process ecosystem. We presented the general modeling and design approach to realize the process infrastructure's architecture transition and provide insights into the modeling process. The realized concepts and the supporting tool represent a proof-of-concept, which has been evaluated with three experts. This evaluation suggests our approach suitable to reach the defined goal and, therefore, motivates further research.

Future work thus includes the stepwise transition of the whole V-Modell~XT infrastructure of which two major components are left: process authoring and the process' build environment. As previous research \cite{DBLP:journals/scp/KuhrmannKT14} has shown, the chosen modeling approach can be transferred to various process metamodels and, therefore, the inclusion of further metamodels and further software processes is possible. Future work also includes the analysis of, e.g., norms and standards to be included in the new ecosystem to provide support for software and systems development in regulated domains.



%

\end{document}